\documentclass[12pt]{article}
\usepackage{graphicx}
\usepackage{multirow}%multirow
\usepackage{booktabs}
\usepackage{array}
\usepackage{booktabs,subcaption,amsfonts,dcolumn}
\usepackage{soul}
\usepackage{authblk} % Manages authors' affiliation
\usepackage{natbib} %comment out if you do not have the package
\usepackage{url} % not crucial - just used below for the URL 
\usepackage{amsmath}
\newcommand{\blind}{0}
\usepackage{enumitem}
\usepackage[doublespacing]{setspace}
\usepackage{bm}
\usepackage{amssymb}
\usepackage[font=scriptsize]{caption} % To reduce caption font size

\DeclareMathOperator{\Var}{Var}
\DeclareMathOperator{\Cov}{Cov}
\DeclareMathOperator{\E}{E}

\newcolumntype{M}[1]{>{\centering\arraybackslash}m{#1}}
\usepackage{booktabs}

\makeatletter
\newcommand*{\addFileDependency}[1]{% argument=file name and extension
  \typeout{(#1)}
  \@addtofilelist{#1}
  \IfFileExists{#1}{}{\typeout{No file #1.}}
}
\makeatother

\newcommand*{\myexternaldocument}[1]{%
    \externaldocument{#1}%
    \addFileDependency{#1.tex}%
    \addFileDependency{#1.aux}%
}

\usepackage{xr} % To get references to supplementary mat in different file
\myexternaldocument{supplementary_materials}

\begin{document}

\def\spacingset#1{\renewcommand{\baselinestretch}%
{#1}\small\normalsize} \spacingset{1}

%%%%%%%%%%%%%%%%%%%%%%%%%%%%%%%%%%%%%%%%%%%%%%%%%%%%%%%%%%%%%%%%%%%%%%%%%%%%%%

\if0\blind
{
  \title{\bf Functional Mixture Regression Control Chart}
   
\author[]{Christian Capezza}
\author[]{Fabio Centofanti}
\author[]{Davide Forcina}
\author[]{Antonio Lepore}
\author[]{Biagio Palumbo\thanks{Corresponding author. e-mail: \texttt{biagio.palumbo@unina.it}}}

\affil[]{Department of Industrial Engineering, University of Naples Federico II, Piazzale Tecchio 80, 80125, Naples, Italy}

\setcounter{Maxaffil}{0}
\renewcommand\Affilfont{\itshape\small}
\date{}
\maketitle

} \fi

\if1\blind
{
  \begin{center}
    {\LARGE\bf Functional Mixture Regression Control Chart}
\end{center}
} \fi

\begin{abstract}
Industrial applications often exhibit multiple in-control patterns due to varying operating conditions, which makes a single functional linear model (FLM) inadequate to capture the complexity of the true relationship between a functional quality characteristic and covariates, which gives rise to the multimode profile monitoring problem.
This issue is clearly illustrated in the resistance spot welding (RSW) process in the automotive industry, where different operating conditions lead to multiple in-control states. 
In these states, factors such as electrode tip wear and dressing may influence the functional quality characteristic differently, resulting in distinct FLMs across subpopulations.
To address this problem, this article introduces the functional mixture regression control chart (FMRCC) to monitor functional quality characteristics with multiple in-control patterns and covariate information, modeled using a mixture of FLMs. 
A monitoring strategy based on the likelihood ratio test is proposed to monitor any deviation from the estimated in-control heterogeneous population. 
An extensive Monte Carlo simulation study is performed to compare the FMRCC with competing monitoring schemes that have already appeared in the literature, and a case study in the monitoring of an RSW process in the automotive industry, which motivated this research, illustrates its practical applicability.
\end{abstract}

\noindent%

\spacingset{1.45} % DON'T change the spacing!

\section{Introduction}
\label{sec_intro}

In modern statistical process monitoring (SPM) applications, experimental measurements of the quality characteristic of interest are often characterized by complex and high-dimensional formats that can best be represented by functional data or profiles \citep{ramsay,kokoszka} and stimulated growing interest in profile monitoring \citep{noorossana2011statistical,zou2012lasso,chou2014monitoring,grasso2014profile,paynabar2016change,wang2018thresholded,ren2019phase,capezza2024robust,capezza2023funcharts}.
Other relevant contributions include those by \cite{jin1999feature}, \cite{colosimo2010comparison}, \cite{grasso2016using}, \cite{grasso2017phase}, \cite{menafoglio2018profile}, \cite{maleki2018overview}, and \cite{jones2021practitioners}.

Profile monitoring, and in general SPM, is commonly implemented in two phases. The first (Phase I) aims to identify a clean dataset to be assumed to be representative of the in-control (IC) state of the process, hereinafter referred to as Phase I or reference sample, through retrospective monitoring of an initial dataset drawn from the process; the second (Phase II) is concerned with prospective monitoring of new observations \citep{qiu2013introduction}, hereinafter referred to as Phase II observations or Phase II dataset.

Traditional profile monitoring methods assume that the IC process is \textit{unimodal}, i.e., it exhibits a single IC state. However, real industrial processes are often \textit{multimodal}, i.e., they operate under different unobserved IC states, and give rise to the so-called \textit{multimode profile monitoring} problem.
To face this problem, \cite{grasso2017phase} propose a method based on curve classification to determine the mode of the data on top of a functional control charting scheme.
\cite{park2014multimode} presented a procedure to monitor time-correlated multimode processes using a mixture of time-series models in a Bayesian framework.
\cite{wang2018registration} developed a feature extraction approach to group and monitor near-circular shape profiles through a likelihood ratio test-based EWMA control chart.

\noindent However, all works on multimode profile monitoring that appeared so far were not designed to incorporate any information from additional concurrent variables, hereinafter referred to also as predictors or covariates, influencing the functional quality characteristic to be monitored, hereinafter referred to also as the response.
When available, incorporating covariate information may drastically improve any SPM scheme's performance.

\noindent It is well known indeed that, if a covariate manifests itself with extreme realization, an SPM scheme based on the observations of the quality characteristic alone may wrongly judge the process out of control (OC) or, dually, there are situations where the covariates are not extreme, and the process may incorrectly appear IC. 

In the profile monitoring framework, a basic solution was given by \cite{centofanti2021bfunctional}, who translated the regression control chart \citep{mandel1969regression} into a functional regression control chart (FRCC) framework and implemented it under the assumption that a functional response is influenced by multiple functional covariates through a functional linear model (FLM). The residuals of the model were then monitored by simultaneously applying Hotelling's $T^2$ and squared prediction error (SPE) control charts, as in \cite{woodall2004using}.
A software implementation of the FRCC is openly available in the R package \texttt{funcharts} \citep{capezza2023funcharts}.
However, while the FRCC is presented as a broader framework, the implementation presented in the original work relies on the assumption that all subjects sampled from the process obey the same FLM, and thus cannot apply in the multimode profile monitoring problem, where subjects belong to more than one subpopulation, in which the FLMs describing the influence of the covariates on the response are different.

In fact, this is the case for the resistance spot welding (RSW) process in the automotive industry that motivated this research.
From a technological perspective, the quality of the RSW process, as also noted by \cite{capezza2021functional}, can be monitored through observations of the dynamic resistance curve (DRC), which is recognized as the quality footprint of the metallurgical development of a spot weld \citep{dickinson1980characterization} and represents the functional quality characteristic of interest, also called the response.
Details on DRC behavior can be found in Section \ref{sec_real}.

The RSW dataset consists of 1802 IC observations, measured in [$m\Omega$], referring to the same spot weld location on different car bodies and 37 OC DRC observations corresponding to known defective spot welds.
\begin{figure}
    \centering
        \includegraphics[width=1\columnwidth]{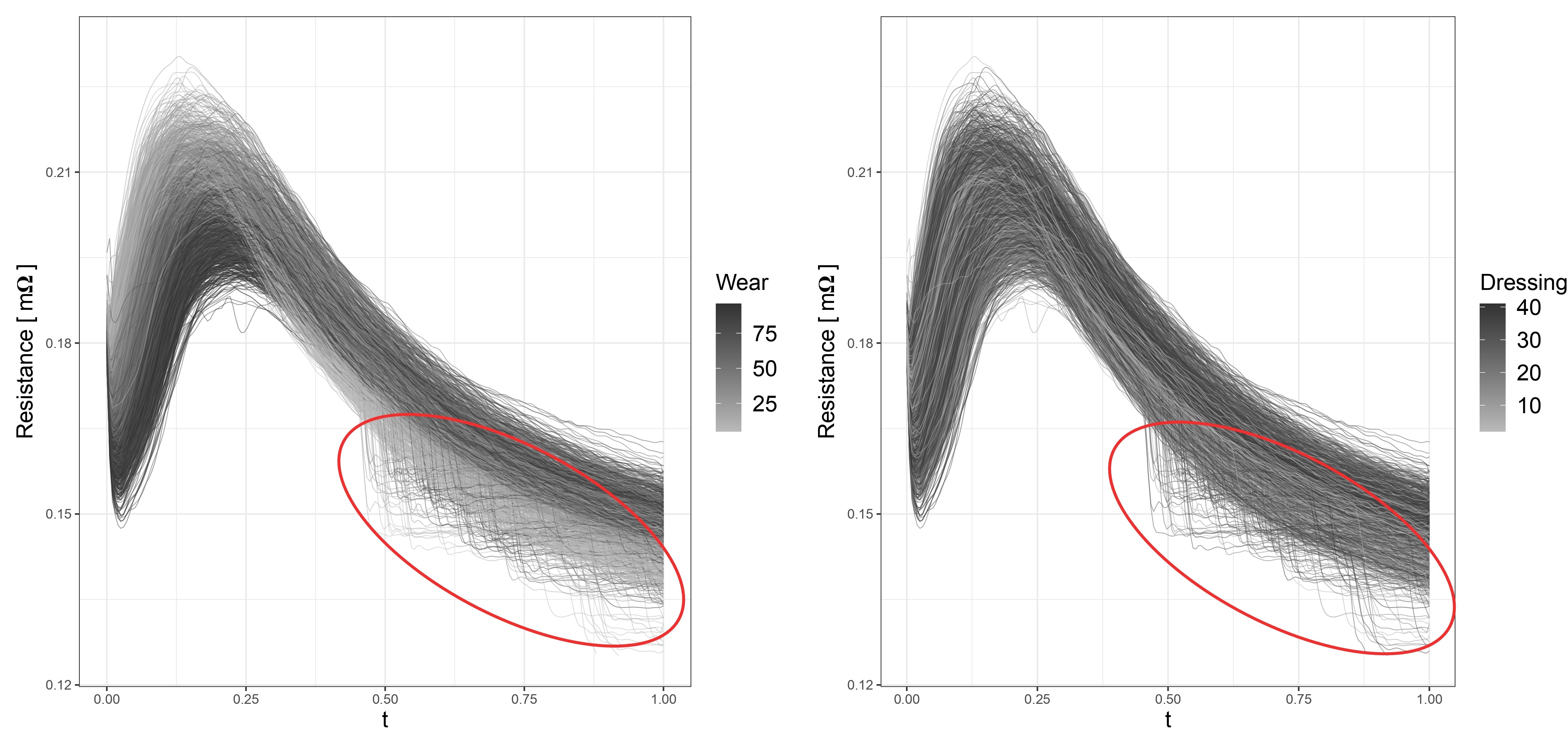}
    \caption{Wear effect (left panel) and dressing effect (right panel) on the DRCs for point 20268. 
    }   
    \label{fig: Wear Dressing effect}
\end{figure}
For illustrative purposes, the left panel of Figure \ref{fig: Wear Dressing effect} displays the IC observations colored by the wear of the electrode tips that performed the spot weld, measured through the number of welds before the electrode tip dressing, whereas, on the right panel, DRCs are colored by the number of electrode tip dressings itself.
These panels clearly show that both the electrode tip wear and the dressing represent informative scalar covariates, which may explain the heterogeneous DRC shapes depicted in Figure \ref{fig: Wear Dressing effect} while not necessarily influencing the final quality of the welded joint.
In addition to these two covariates, many other phenomena can cause DRC heterogeneity across different subpopulations. One known phenomenon is the expulsion of molten material caused by an abnormal welding current flowing through the electrode tips due to an abnormal clamping force that causes a significant drop in the DRCs and may cause the DRC behavior highlighted by the balloons in the left and right panels of Figure \ref{fig: Wear Dressing effect}.

Mixtures of regression models are undoubtedly useful for modeling such heterogeneity, as pointed out by \citep{desarbo1988maximum,jones1992fitting} in the case of scalar response versus scalar predictor variables.
Extensions to the functional case are presented by \cite{yao2011functional,ciarleglio2016wavelet,zhao2012wavelet} in the scalar-on-function regression setting. In particular, \cite{yao2011functional} proposed a mixture of functional regression models which relate a univariate scalar response with a functional predictor variable by adopting the eigenbasis representation of the latter, reducing to a framework similar to the classic mixture of regression models.
\cite{wang2016mixture} studied the mixture of functional regression models in the case of a univariate functional response and multivariate functional covariates through concurrent FLMs.
Finally, \cite{devijver2017model} extended 
this case with one functional predictor variable to non-concurrent FLMs, also referred to as total FLMs \citep{ramsay}, where the value of a univariate functional response is related to the entire domain of a functional predictor variable.
It is important to note that all functional mixture regression (FMR) models are more flexible in modeling unobserved heterogeneity than existing functional clustering methods. The latter methods focus only on clustering the functions themselves, while the former focuses on detecting clusters characterized by different regression structures.

In this article, a functional mixture regression control chart (FMRCC) is proposed to deal with the multimodal profile monitoring problem through an FMR approach with functional response and multivariate functional or scalar predictors instead of modeling different modes only through functional clustering (see, e.g., \cite{capezza2021functional,centofanti2024sparse}). To deal with the infinite dimensionality of the data, both the response and predictors are projected onto an appropriate basis function system. This allows for the application of existing methods based on the finite mixture regression model and the expectation-maximization (EM) algorithm on the basis coefficients. 
Finally, an SPM strategy based on the likelihood ratio test (LRT) is proposed to monitor any deviation from the estimated IC heterogeneous population. 
A Monte Carlo simulation study is performed to assess the SPM performance of the FMRCC scheme. 
The practical applicability of the FMRCC is demonstrated through a  case study in the monitoring of the aforementioned RSW process.

The remainder of the article is organized as follows. Section \ref{methodology} provides a complete description of the FMRCC. In Section \ref{sec_sim}, the performance of the FMRCC is compared to that of the other three control charts by means of a simulation study. In Section \ref{sec_real}, the case study is presented. Section \ref{sec_conclusions} concludes the article.
The Supplementary Materials contains additional details on the data generation process in the simulation study.
All calculations and plots were obtained using the programming language R \citep{r2021}.

\section{The Functional Mixture Regression Control Chart Framework}
\label{methodology}
\subsection{Preliminaries}
\label{preliminaries}
The proposed FMRCC is a general framework for multimode profile monitoring where different modes are characterized by different FLMs, with a univariate functional response. 
Assume that $N$ observations of random functions $\tilde{X_{1}},\dots,\tilde{X_{p}}$ and $\tilde{Y}$ are available with $\tilde{X_{j}}$, $j = 1,\dots,p$, representing predictors having values in $L^2(\mathcal{S})$, and $\tilde{Y}$ representing the response having values in $L^2(\mathcal{T})$.
$L^2(\mathcal{D})$ denotes the Hilbert space of square-integrable functions defined in the compact set $\mathcal{D} \subseteq \mathbb{R}$, with the inner product $\langle f,g\rangle = \int_{\mathcal{S}}f(s)g(s)ds$.
Let $\mathbb{H}^{\Tilde{X}} = (L^2(\mathcal{S}))^p$ denote the Hilbert space whose elements are vectors of functions in $L^2(\mathcal{S})$. 
Then $\Tilde{\boldsymbol{X}} = (\tilde{X_1},\dots,\Tilde{X_p})^T$ is a random vector of functions, whose realizations are in $\mathbb{H}^{\Tilde{X}}$.
Assume that $\boldsymbol{\Tilde{X}}$ has a smooth mean function $\boldsymbol{\mu}^{\Tilde{X}} = (\mu_1^{\Tilde{X}},\dots,\mu_p^{\Tilde{X}})^T$, with $\mu_j^{\Tilde{X}} = \E(\Tilde{X}_j)$ and a covariance function $\boldsymbol{C}^{\Tilde{X}} = \{C_{i,j}^{\Tilde{X}}\}_{1\le {i,j}\le p}$, with $C_{i,j}^{\tilde{X}}(s_1,s_2) = \Cov(\Tilde{X}_i(s_1),\Tilde{X}_j(s_2))$, for $s_1,s_2 \in \mathcal{S}$. Analogously, let ${\mu}^{\Tilde{Y}} = \E(\Tilde{Y})$ and $C^{\Tilde{Y}}(t_1,t_2) = \Cov(\tilde{Y}(t_1),\tilde{Y}(t_2))$, for $t_1,t_2 \in \mathcal{T}$, be the mean and covariance functions of the response variable, respectively.

\noindent In particular, to avoid exhibiting incomparable magnitudes of variation, the transformation approach of \cite{chiou2014multivariate} is utilized hereinafter.
To this end, let $\boldsymbol{X} = (X_1,\dots,X_p)^T = (\boldsymbol{V}^{\tilde{X}})^{-1} (\boldsymbol{\Tilde{X}} - \boldsymbol{\mu}^{\tilde{X}})$ be the standardized covariate with $\boldsymbol{V}^{\Tilde{X}} = \text{diag}((v_1^{\Tilde{X}})^{1/2},\dots,(v_p^{\Tilde{X}})^{1/2})$ and the square root of $v_j^{\Tilde{X}}(s) = {C}_{j,j}^{\Tilde{X}}(s,s)$, $\forall s \in \mathcal{S}$.
Consequently, the standardized response is defined as $Y = ({v^{\tilde{Y}}})^{-1/2} ({\Tilde{Y}} - {\mu}^{\tilde{Y}})$ with $v^{\Tilde{Y}}(t) = {C}^{\Tilde{Y}}(t,t)$, $\forall t \in \mathcal{T}$.
Then, let $\boldsymbol{C}^X$ and $C^Y$ denote the covariance functions of the standardized covariates and the response.
There exists a multivariate orthonormal eigenbasis $\{ \boldsymbol{\psi}_l^X(s) \}_{l=1,2,\dots}$, such that $\langle\boldsymbol{\psi}_l^X,\boldsymbol{\psi}_h^X \rangle_{\mathbb{H}^X} = \delta_{lh}$, with $\delta_{lh}$ the Kronecker delta. The corresponding non-negative eigenvalues $\{\lambda_l^X\}_{l=1,2,\dots}$ are nonnegative and are supposed, without loss of generality, to be nonincreasing with $l$ and such that $\boldsymbol{C}^X(s_1,s_2) = \sum_{l = 1}^\infty \lambda_l^X \boldsymbol{\psi}_l^X(s_1) \boldsymbol{\psi}_l^X(s_2)^T$, for $s_1,s_2 \in \mathcal{S}$.
Similarly, there exists an orthonormal eigenbasis $\{\psi_m^Y(t) \}_{m=1,2,\dots}$ such that $C^Y(t_1,t_2) = \sum_{m = 1}^\infty \lambda_m^Y \psi_m^Y(t_1) \psi_m^Y(t_2)^T$, for $t_1,t_2 \in \mathcal{T}$ where $\{\lambda_m^Y\}_{m=1,2,\dots}$ are the corresponding nondecreasing eigenvalues such that $\langle\psi_l^Y,\psi_h^Y \rangle_{\mathbb{H}^X} = \delta_{lh}$.

\subsection{Model specification}
Suppose $N$ observations are spread among an unknown number $K$ of mutually exclusive clusters, each of which is characterized by a different FLM describing the influence of the covariates on the response, with the probability that each observation belonging to the $k$-th cluster is $\pi_k$, with $\sum_{k = 1}^K\pi_k = 1$. Moreover, let us denote by $\bm Z = (Z_{1},\dots,Z_{K})^T$, $k = 1,\dots,K$, the unknown component-label vector corresponding to $(\boldsymbol{\Tilde{X}},\Tilde{Y})$, where $Z_{k}$ equals 1 if the observation is in the $k$-th cluster and 0 otherwise. Conditioned to $Z_{k} = 1$, that is, to the $k$-th cluster, the functional linear model to be estimated is
\begin{equation}
\label{model}
    Y(t) = \beta_{0k}(t) + \int_{\mathcal{S}} (\boldsymbol{\beta}_k(s,t))^T\boldsymbol{X}(s)ds + \varepsilon(t), \quad t \in \mathcal{T}, k = 1,\dots,K,
\end{equation}
where $\boldsymbol{\beta}_k = (\beta_{k1},\dots,\beta_{kp})$ is $k$-th regression coefficient vector, whose elements are bivariate functions in the space of square-integrable functions defined on the closed interval $\mathcal{S}\times\mathcal{T}$, $\beta_{0k}$ and ${\varepsilon}$ are $k$-th  functional intercept and the functional error term, respectively, both defined on the compact domain $\mathcal{T}$. The functional error term $\varepsilon$ is supposed independent of $\boldsymbol{X}$ and having $\E(\varepsilon) = 0$ and $\Var(\varepsilon) = \nu_\varepsilon^2$.
Thus, the regression function for the $k$-th cluster is 
\begin{equation}
\label{FLMk regression fun}
     \E(Y(t)|\boldsymbol{X}) = \beta_{0k}(t) + \int_{\mathcal{S}} (\boldsymbol{\beta}_k(s,t))^T\boldsymbol{X}(s)ds, \quad t \in \mathcal{T}, k = 1,\dots,K.
\end{equation}
From the multivariate functional principal component (MFPC) decomposition \citep{chiou2014multivariate,happ2018multivariate}, the predictors and the response can be represented as 
\begin{equation}
\label{decomposition}
     \boldsymbol{X}(s) = \sum_{l=1}^\infty \xi_l^X \boldsymbol{\psi}_l^X(s),\quad Y(t) = \sum_{m=1}^\infty \xi_m^Y {\psi_m^Y(t)},
\end{equation}
where $\{ \boldsymbol{\psi}_l^X(s) \}$ and $\{ {\psi}_m^Y(t) \}$ are the eigenfunctions defined in section \ref{preliminaries} and $\xi_l^X = {\langle \boldsymbol{X},\boldsymbol{\psi}_l^X \rangle}_{\mathbb{H}^X}$ and $\xi_m^Y = {\langle {Y},{\psi_m}^Y \rangle}$ are the functional principal component scores which, unconditionally, are uncorrelated and satisfy $\E(\xi_l^X) = 0$, $\E(\xi_l^X\xi_h^X) = \lambda_l^X\delta_{lh}$ and $\E(\xi_m^Y) = 0$, $\E(\xi_m^Y\xi_h^Y) = \lambda_m^Y\delta_{mh}$.
Since the eigenfunctions of a square-integrable random function form a complete orthonormal basis, the regression coefficient vector and the functional intercept can be expressed as
\begin{multline}
    \label{beta expansion}
     \boldsymbol{\beta}_k(s,t) = \sum_{l,m=1}^\infty b_{lmk}\boldsymbol{\psi}_l^X(s)\psi_m^Y(t), \quad {\beta_{0k}(t)} = \sum_{m=1}^\infty b_{0mk}\psi_m^Y(t), \quad s \in \mathcal{S}, t \in \mathcal{T},k = 1,\dots,K,
\end{multline}
respectively, where $b_{lmk}$ and $b_{0k}$ are the coefficents of the basis expansions. 
Furthermore, considering the following expansion of the functional error term $\varepsilon$,
\begin{equation}
\label{eps}
    \varepsilon(t) = \sum_{m=1}^\infty \varepsilon_m \psi_m^Y(t),
\end{equation}
with $\varepsilon_m = \langle \varepsilon,\psi_m^Y \rangle$, and by plugging Equations \eqref{decomposition}, \eqref{eps} and \eqref{beta expansion} in \eqref{model}, we obtain
\begin{multline}
    \label{model after substitution}
    Y(t) =  \sum_{m=1}^\infty b_{0mk}\psi_m^Y(t)\,+\, \int_{\mathcal{S}} \left( \sum_{l,m=1}^\infty b_{lmk}\boldsymbol{\psi}_l^X(s)\psi_m^Y(t) \right)^T \sum_{l=1}^\infty \xi_l^X \boldsymbol{\psi}_l^X(s)ds
    \,+\, \sum_{m=1}^\infty \varepsilon_m \psi_m^Y(t),\\
    \quad t \in \mathcal{T},k = 1,\dots,K.
\end{multline}
In the context of multivariate functional regression, the estimation of model parameters in Equation \eqref{model after substitution} requires regularization of the predictor and response functions, which is achieved by considering the truncated principal component decomposition $\boldsymbol{X}_L$ and $Y_M$ of $\boldsymbol{X}$ and $Y$. That is,
\begin{equation}
\label{truncated}
    \boldsymbol{X}_L(s) = \sum_{l=1}^L \xi_l^X \boldsymbol{\psi}_l^X(s),
    \quad
    Y_M(t) = \sum_{m=1}^M \xi_m^Y \psi_m^Y(t),
\end{equation}
where $L$ and $M$ denote the minimum number of components necessary to explain a given fraction of the total variation in the data.
Other selection criteria, e.g., based on Akaike information criterion (AIC) or Bayesian information criterion (BIC), could also be used.
Accordingly, we get the truncated version of Equation \eqref{beta expansion} as follows
\begin{multline}
        \label{beta expansion truncated}
         \boldsymbol{\beta}_k(s,t) = \sum_{l=1}^L\sum_{m=1}^M b_{lmk}\boldsymbol{\psi}_l^X(s)\psi_m^Y(t), \quad {\beta_{0k}(t)} = \sum_{m=1}^M b_{0mk}\psi_m^Y(t), \quad s \in \mathcal{S}, t \in \mathcal{T},k = 1,\dots,K.
\end{multline}
Then, using $Y_M(t)$ in Equation \eqref{truncated} instead of $Y(t)$ and by the orthonormality of $\{ \boldsymbol{\psi}_l^X(s) \}$ and $\{{\psi}_m^Y(t) \}$, Equation \eqref{model} is reduced to the truncated multivariate linear regression model
\begin{equation}
    \label{multivariate linear regression intercept}
    \boldsymbol{\xi}^Y_M = \boldsymbol{b}_{0k} + (\boldsymbol{B}_{LMk})^T \boldsymbol{\xi}^X_L + \boldsymbol{\varepsilon}_M,\quad k = 1,\dots,K,
\end{equation}
where $\boldsymbol{\xi}_M^Y = (\xi_1^Y,\dots,\xi_M^Y)^T$, $\boldsymbol{\xi}_L^X = (\xi_1^X,\dots,\xi_L^X)^T$, $\boldsymbol{\varepsilon}_M$ is the truncated version of the basis coefficients of $\varepsilon$ in Equation \eqref{eps}, that is, multivariate Gaussian random errors with zero mean and covariance $\boldsymbol{\Sigma}_k$, $\boldsymbol{B}_{LMk} = \{b_{lmk}\}_{l =1,\dots, L,m=1,\dots, M}$ and $\boldsymbol{b}_{0k} = (b_{01k},\dots,b_{0Mk})$.
Additionally, Equation \eqref{multivariate linear regression intercept} can be reduced to 
\begin{equation}
    \label{multivariate linear regression}
    \boldsymbol{\xi}^Y_M = (\boldsymbol{B}_{LMk})^T \boldsymbol{\xi}^X_L + \boldsymbol{\varepsilon}_M,\quad k = 1,\dots,K,
\end{equation}
by incorporating the intercept term $\boldsymbol{b}_{0k}$ in $\boldsymbol{B_{LMk}}$ and by adding 1 to the head of $\boldsymbol{\xi}^X_L$.
Hence, given $N$ independent realizations $(\boldsymbol{X}_i,Y_i)$ of $(\boldsymbol{X},Y)$, $i = 1,\dots,N$, the probability density function (pdf) of $\boldsymbol{\xi}_{i,M}^Y|\boldsymbol{\xi}_{i,L}^X$ comes from the following mixture:
\begin{equation}
    \label{mixture density}
    f(\boldsymbol{\xi}_{i,M}^Y|\boldsymbol{\xi}_{i,L}^X) = \sum_{k=1}^K \pi_k \phi(\boldsymbol{\xi}_{i,M}^Y; (\boldsymbol{B}_{LMk})^T \boldsymbol{\xi}^X_{i,L}, \boldsymbol{\Sigma}_k),
\end{equation}
where $\phi(\cdot,\boldsymbol{\mu},\boldsymbol{\Sigma})$ is the multivariate Gaussian pdf with mean $\boldsymbol{\mu}$ and covariance $\boldsymbol{\Sigma}$.
Equation \eqref{mixture density} is the classical $K$-component Gaussian mixture regression model \citep{mclachlan2004finite}.
Therefore, the log-likelihood function corresponding to $(\boldsymbol{\xi}_{1,M}^Y,\boldsymbol{\xi}_{1,L}^X),\dots,(\boldsymbol{\xi}_{N,M}^Y,\boldsymbol{\xi}_{N,L}^X)$ is given by
\begin{equation}
    \label{log-likelihood}
    \log L(\boldsymbol{\Psi}) = \sum_{i = 1}^N \log \sum_{k=1}^K \pi_k \phi(\boldsymbol{\xi}_{i,M}^Y;(\boldsymbol{B}_{LMk})^T \boldsymbol{\xi}^X_{i,L}, \boldsymbol{\Sigma}_k),
\end{equation}
where $\boldsymbol{\Psi} = (\pi_k, \boldsymbol{B}_{LMk}, \boldsymbol{\Sigma}_k)_{k=1,\dots,K}$ is the parameter set to be estimated.
In the presence of a large number of components $K$, having a different matrix parameter $\boldsymbol{\Sigma}_k$ in each component may lead to too many parameters and then to overfitting. 
Therefore, we also consider alternative and more parsimonious parameterizations of $\boldsymbol{\Sigma}_k$.
Then, the problem of model selection is discussed at the end of the next Section \ref{sec_estimation}.
In particular, in this article we consider the following parametrizations, i.e., the spherical family with one common scalar parameter $\lambda$ and diagonal matrix $\boldsymbol{\Sigma}_k = \lambda\boldsymbol{I}$ for each $k=1,\dots, K$; the spherical family with $K$ scalar parameters $\lambda_1, \dots, \lambda_K$, where $\boldsymbol{\Sigma}_k = \lambda_k\boldsymbol{I}$; the parameterization with common covariance $\boldsymbol{\Sigma}_k = \boldsymbol{\Sigma}$, not necessarily diagonal; the full parametrization where each component $k=1,\dots, K$ has its own covariance $\boldsymbol{\Sigma}_k$.

\subsection{The estimation method}
\label{sec_estimation}
An estimator $\hat{\boldsymbol{\Psi}}$ of $\boldsymbol{\Psi}$ in Equation \eqref{log-likelihood} can be calculated by maximizing the log-likelihood in Equation \eqref{log-likelihood}.
Unfortunately, maximization should be performed numerically through dedicated algorithms, such as the expectation-maximization (EM) algorithm \citep{dempster1977maximum,mclachlan2007algorithm}.
The EM algorithm first requires the construction of the complete-data log-likelihood by elaborating Equation \eqref{log-likelihood} as follows
\begin{equation}
\label{complete-data log-likelihood}
\log L_c(\boldsymbol{\Psi}) = \sum_{i=1}^N\sum_{k=1}^K Z_{ki} \log [\pi_k \phi(\boldsymbol{\xi}_{i,M}^Y;(\boldsymbol{B}_{LMk})^T \boldsymbol{\xi}^X_{i,L}, \boldsymbol{\Sigma}_k)].
\end{equation}
Then, starting from an initial solution $\boldsymbol{\Psi}^{(0)}$, the version of the EM algorithm for a mixture regression model \citep{jones1992fitting} such as that introduced in Equation \eqref{mixture density} alternates between the expectation and maximization steps, referred to as the E step and M step, respectively, until there is no appreciable change in the logarithmic likelihood values.
At iteration $q$, the E-step computes the conditional expectation of the complete-data log-likelihood function in Equation \eqref{complete-data log-likelihood}, using the current parameter vector $\boldsymbol{\Psi}^{(q)}$; that is
\begin{equation}
    \label{e-step}
    Q(\boldsymbol{\Psi},\boldsymbol{\Psi}^{(q)}) = \sum_{i=1}^N\sum_{k=1}^K \tau_{ki}^{(q)}\log[\pi_k \phi(\boldsymbol{\xi}_{i,M}^Y; (\boldsymbol{B}_{LMk})^T \boldsymbol{\xi}^X_{i,L}, \boldsymbol{\Sigma}_k)].
\end{equation}
The E-step only requires computing the posterior probabilities of component membership $\tau_{ki}^{(q)}$, $i=1,\dots,N$, for each of the $K$ components, namely
\begin{equation}
    \label{posterior}
    \tau_{ki}^{(q)} =\frac{\pi_k^{(q)} \phi(\boldsymbol{\xi}_{i,M}^Y;  (\boldsymbol{B}_{LMk}^{(q)})^T \boldsymbol{\xi}^X_{i,L}, \boldsymbol{\Sigma}_k^{(q)})}{\sum_{h=1}^K\pi_h^{(q)} \phi(\boldsymbol{\xi}_{i,M}^Y; (\boldsymbol{B}_{LMh}^{(q)})^T \boldsymbol{\xi}^X_{i,L}, \boldsymbol{\Sigma}_h^{(q)})}.
\end{equation}
The M-step updates the value of the parameter vector $\boldsymbol{\Psi}$ by maximizing Equation \eqref{e-step} with respect to $\boldsymbol{\Psi}$.
The mixing proportions updates are
\begin{equation}
    \label{mixing updates}
    \pi_k^{(q+1)} = \frac{1}{N}\sum_{i=1}^N \tau_{ki}^{(q)},
\end{equation}
while the regression parameters $\boldsymbol{B}_{LMk}$ and $\boldsymbol{\Sigma}_k$ are obtained by maximizing the complete-data log-likelihood function in Equation \eqref{e-step} with respect to $(\boldsymbol{B}_{LMk}, \boldsymbol{\Sigma}_k)$.
This corresponds to solving $K$ least-squares problems with the posterior probabilities $(\tau_{ki}^{(q)})_{k=1,\dots,K}$ as weights, which give
\begin{equation}
    \label{beta}
    \boldsymbol{B}_{LMk}^{(q+1)} = \left[ \sum_{i=1}^N \tau_{ki}^{(q)}(\boldsymbol{\xi}_{i,L}^X)^T\boldsymbol{\xi}_{i,L}^X \right]^{-1} \sum_{i=1}^N\tau_{ki}^{(q)}\boldsymbol{\xi}_{i,L}^X (\boldsymbol{\xi}_{i,M}^Y)^T,
\end{equation}
\begin{equation}
    \label{sigma}
    \boldsymbol{\Sigma}_{k}^{(q+1)} = \frac{1}{\sum_{i=1}^N\tau_{ki}^{(q)}}
    \sum_{i=1}^N \tau_{ki}^{(q)} \left[\boldsymbol{\xi}_{i,M}^Y - (\boldsymbol{B}_{LMk}^{(q+1)})^T\boldsymbol{\xi}_{i,L}^X\right]\left[\boldsymbol{\xi}_{i,M}^Y - (\boldsymbol{B}_{LMk}^{(q+1)})^T\boldsymbol{\xi}_{i,L}^X\right]^T.
\end{equation}
\cite{dempster1977maximum} showed that at each iteration of the EM algorithm, the log-likelihood in Equation \eqref{log-likelihood} is non-decreasing. Additionally, the series of parameter estimates converges towards a local maximum of the log-likelihood function \citep{wu1983convergence}.\\
As suggested in \cite{mclachlan2004finite} and implemented in popular software packages \texttt{EMMIXskew} \citep{wang2009multivariate} and \texttt{mclust} \citep{scrucca2016mclust}, the K-means algorithm can be utilized to obtain an initial solution $\boldsymbol{\Psi}^{(0)}$.
The BIC is used for the parameterization of the covariance matrix $\boldsymbol{\Sigma}_k$ as well as for the choice of the number of clusters $K$, when the number of components is not known prior to or evident by data exploration.
BIC is one of the most widely adopted methods and has provided good results in various applications of model-based clustering \citep{fraley2002model}.

\subsection{The SPM strategy}
\label{monitoring}
Let us assume that a set of functional principal component scores $(\boldsymbol{\xi}_{i,L}^X,\boldsymbol{\xi}_{i,M}^Y)$ corresponding to the functional observations $(\boldsymbol{\Tilde{X}}_i,\Tilde{Y}_i)$, $i = 1,\dots,N$, is available from Phase I, and that it is drawn from an IC heterogeneous population as defined by \cite{jones1992fitting} and denoted by $\boldsymbol{\Pi}$. Then, let us suppose that we are interested in testing whether the scores $(\boldsymbol{\xi}_{N+1,L}^X,\boldsymbol{\xi}_{N+1,M}^Y)$ of a new functional observation
$(\boldsymbol{\Tilde{X}}_{N+1},\Tilde{Y}_{N+1})$ belong to
$\boldsymbol{\Pi}$ or not.
That is, we are interested in testing the following null hypothesis $H_0$ (IC process) versus the alternative hypothesis $H_1$ (OC process)
\begin{equation}
    \label{H0}
    H_0: \boldsymbol{\xi}_{N+1,M}^Y|\boldsymbol{\xi}_{N+1,L}^X \in \boldsymbol{\Pi}, \quad
    % \label{H1}
    H_1: \boldsymbol{\xi}_{N+1,M}^Y|\boldsymbol{\xi}_{N+1,L}^X \notin \boldsymbol{\Pi}.
\end{equation}
The test statistic can be constructed using a likelihood ratio test (LRT), as the ratio between the likelihood functions maximized under $H_0$ and $H_1$, and denoted by $L_0(\boldsymbol{\Psi})$ and $ L_1(\boldsymbol{\Psi},\boldsymbol{\beta})$, respectively.
The former can be written as
\begin{equation}
    \label{likelihood H0}
    L_0(\boldsymbol{\Psi}) = \left(\prod_{i=1}^N f(\boldsymbol{\xi}_{i,M}^Y|\boldsymbol{\xi}_{i,L}^X;\boldsymbol{\Psi})\right)f(\boldsymbol{\xi}_{N+1,M}^Y|\boldsymbol{\xi}_{N+1,L}^X;\boldsymbol{\Psi}),
\end{equation}
where $f(\boldsymbol{\xi}_{N+1,M}^Y|\boldsymbol{\xi}_{N+1,L}^X;\boldsymbol{\Psi})$ is the IC pdf defined as in Equation \eqref{mixture density} evaluated at a new observation,
whereas the latter is similarly expressed as
\begin{equation}
    \label{likelihood H1}
    L_1(\boldsymbol{\Psi},\boldsymbol{\beta}) = \left(\prod_{i=1}^N f(\boldsymbol{\xi}_{i,M}^Y|\boldsymbol{\xi}_{i,L}^X;\boldsymbol{\Psi})\right)h(\boldsymbol{\xi}_{N+1,M}^Y|\boldsymbol{\xi}_{N+1,L}^X;\boldsymbol{\beta}),
\end{equation}
where $h(\boldsymbol{\xi}_{i, M}^Y|\boldsymbol{\xi}_{i, L}^X;{\boldsymbol{\beta}})$ denotes the pdf with parameter vector $\boldsymbol{\beta}$ from which the new observation is sampled.
However, since there is only one observation under $H_1$, according to \cite{wang1997new}, a simple constant density $h(\boldsymbol{\xi}_{N+1, M}^Y|\boldsymbol{\xi}_{N+1, L}^X)\equiv c$ should be used and dropped from the LRT statistic in these cases. Therefore, following the principle of equal ignorance, the outlier distribution is not explicitly required \citep{sain1999outlier}.
By denoting the likelihood based on the Phase I sample with
\begin{equation}
    \label{L_tilde_1}
    \Tilde{L}_1(\boldsymbol{\Psi}) = \prod_{i=1}^N f(\boldsymbol{\xi}_{i,M}^Y|\boldsymbol{\xi}_{i,L}^X;\boldsymbol{\Psi}),
\end{equation}
the LRT statistic can be written as 
\begin{equation}
    \label{lrt statistic}
    \Lambda = \frac{\underset{\boldsymbol{\Psi}\in \boldsymbol{\Omega}}{\sup}L_0(\boldsymbol{\Psi})}{\underset{\boldsymbol{\Psi}\in \boldsymbol{\Omega}}{\sup} \Tilde{L_1}(\boldsymbol{\Psi})},
\end{equation}
where $\boldsymbol{\Omega}$ is the complete parameter space.
In particular, the denominator of Equation \eqref{lrt statistic} is computed only once by plugging in the estimate $\boldsymbol{\hat{\Psi}}$ of the parameter set $\boldsymbol{\Psi}$ calculated, as described in Section \ref{sec_estimation}, with the Phase I sample,
whereas the numerator $\underset{\boldsymbol{\Psi}\in \boldsymbol{\Omega}}{\sup}L_0(\boldsymbol{\Psi})$ is computed at each new observation, using $\boldsymbol{\hat{\Psi}}$ as an initial value. 
\cite{wang1997new} demonstrated that, as $n \to \infty$, $\Lambda$ approaches $f(\boldsymbol{\xi}_{N+1,M}^Y|\boldsymbol{\xi}_{N+1,L}^X;\boldsymbol{\hat{\Psi}})$. That is, for large $n$, due to the minimal conditioning effect that a new observation has on the maximizer of Equation \eqref{likelihood H0}, there is no significant change in the parameter estimate.
This asymptotic approximation, which is used throughout this article, leads to the simplification of Equation \eqref{lrt statistic}, which, for large $n$, can be rewritten as
\begin{equation}
\label{monitor_pre}
    \Lambda = f(\boldsymbol{\xi}_{N+1,M}^Y|\boldsymbol{\xi}_{N+1,L}^X;\boldsymbol{\Hat{\Psi}}).
\end{equation}
For obvious reasons, the monitoring statistic $W = - \log \Lambda$ can be used more practically in place of \eqref{monitor_pre}.\\
Unlike \cite{sain1999outlier} and \cite{wang1997new}, who calculate the distribution of the test statistic under $H_0$ using a bootstrap procedure, here we assume that the Phase I sample is suitably partitioned into a \textit{training} and \textit{tuning} datasets as also done in \cite{colosimo2010comparison}, \cite{capezza2024robust} and \cite{centofanti2021bfunctional}, to reduce the undesirable effect caused by uncertainty in the estimation of the MFPCA model \citep{ramaker2004effect,kruger2012statistical}.
The training set is used to estimate the model parameters, whereas the tuning set is used to estimate the control chart limit for the monitoring statistic calculated based on the estimated model. 
The upper control limit $W^{lim}_\alpha$ is obtained non-parametrically by means of the $(1-\alpha)$-quantile of the empirical distribution of $W$, where $\alpha$ is the desired type-I error rate.\\
In Phase II monitoring, to test hypotheses in Equation \eqref{H0} based on the current functional observation, denoted by $(\boldsymbol{\Tilde{X}}^*,\Tilde{Y}^*)$, the functional principal component scores $\hat{{\xi}}_l^{X*} = \langle \boldsymbol{X}^*,\boldsymbol{\hat{\psi}}_l^X\rangle_{l =1,\dots,L}$ and ${\hat{\xi}}_m^{Y*} = \langle Y^*,\hat{\psi}_m^Y\rangle_{m =1,\dots,M}$ are obtained from $\boldsymbol{X}^* = (\hat{\boldsymbol{V}}^{{\tilde{X}}})^{-1} (\boldsymbol{\Tilde{X}}^* - \hat{\boldsymbol{\mu}}^{\tilde{X}})$ and $Y^*=({\hat{v}^{\tilde{Y}}})^{-1/2} ({\Tilde{Y}^*} - \hat{{\mu}}^{\Tilde{Y}})$.
Then, the corresponding realization of the monitoring statistic denoted by $\hat{W}^*$ is calculated as 
\begin{equation}
    \label{statistic approx}
     \hat{W}^* = - \log f(\hat{\boldsymbol{\xi}}_{M}^{Y*}|\hat{\boldsymbol{\xi}}_{L}^{X*};\boldsymbol{\hat{\Psi}}).
\end{equation}
If $\hat{W}^* > W^{lim}_\alpha$, we reject $H_0$, and an alarm is issued to signal a possible OC state of the process.

\subsubsection{The studentized FMRCC}
To reduce the influence of covariate mean shifts on the residual mean, especially when the Phase I sample size is small, as in \cite{centofanti2021bfunctional}, we propose a \textit{studentized} version of the residuals and thus of the monitoring statistic
\begin{equation}
    \label{stat_stud}
    \hat{W}^* = -\log\sum_{k=1}^K \hat{\pi}_k \phi\left(\hat{\boldsymbol{\xi}}_{M}^{Y*};(\hat{\boldsymbol{B}}_{LMk})^T\hat{\boldsymbol{\xi}}_{L}^{X*},\hat{\boldsymbol{\Sigma}}_{k}^*\right)
\end{equation}
to be used in place of Equation \eqref{statistic approx}.
In Equation \eqref{stat_stud}, the elements of $\hat{\boldsymbol{\Sigma}}_{k}^*$, which is the covariance matrix of the prediction error of $(\boldsymbol{\Tilde{X}}^*,\Tilde{Y}^*)$, are calculated as
\begin{equation}
     \hat{\sigma}_{rhk}^* = \left(\hat{\sigma}_{rhk} + (\hat{\boldsymbol{\xi}}_{L}^{X*})^T \Cov\{\hat{\boldsymbol{B}}_{LM(r)k},\hat{\boldsymbol{B}}_{LM(h)k}\}\hat{\boldsymbol{\xi}}_{L}^{X*}\right)_{r,h = 1,\dots,M},
\end{equation}
where $\hat{\sigma}_{rhk}$ is the element corresponding to the $r$-th row and the $h$-th column of $\hat{\boldsymbol{\Sigma}}_{k}$, $\hat{\boldsymbol{B}}_{LM(r)k}$ and $\hat{\boldsymbol{B}}_{LM(h)k}$ are the $r$-th and $h$-th row of $\hat{\boldsymbol{B}}_{LMk}$, respectively.
Recalling the EM algorithm, $\Cov\{\hat{\boldsymbol{B}}_{LM(r)k},\hat{\boldsymbol{B}}_{LM(h)k}\}$ can be obtained accordingly as
 \begin{equation}
     \Cov\{\hat{\boldsymbol{B}}_{LM(r)k},\hat{\boldsymbol{B}}_{LM(h)k}\} = \hat{\sigma}_{rhk}\left[ \hat{\boldsymbol{\xi}}_{L}^{X}\boldsymbol{\Lambda}_k(\hat{\boldsymbol{\xi}}_{L}^{X})^T\right]^{-1} \hat{\boldsymbol{\xi}}_{L}^{X} \boldsymbol{\Lambda}_k \boldsymbol{\Lambda}_k(\hat{\boldsymbol{\xi}}_{L}^{X})^T \left[ \hat{\boldsymbol{\xi}}_{L}^{X} \boldsymbol{\Lambda}_k (\hat{\boldsymbol{\xi}}_{L}^{X})^T \right]^{-1},
 \end{equation}
 where $\boldsymbol{\Lambda}_k$ is the $n \times n$ diagonal matrix with diagonal elements the posterior probabilities $\tau_{ki}$, $i = 1, \dots N$. In this article, we will always use the studentized FMRCC when not differently specified.
 
\section{Simulation Study}
\label{sec_sim}
\subsection{Data generation}
To evaluate the performance of the proposed FMRCC, the functional response has been generated according to the following FLM
\begin{equation}
\label{data gen}
    Y(t) = (1-\Delta_2) \beta_{0k}(t) + \int_{\mathcal{S}} \Delta_2(\beta_k(s,t))^TX(s)ds + \varepsilon(t), \quad t \in \mathcal{T}, k = 1,\dots,K,
\end{equation}
by setting $K=3$ clusters and $p=1$ functional covariate.
On top of these choices, which do not impact the generality of the study, three different simulation scenarios are generated based on the difference in the FLM structure across clusters. The parameter $\Delta_2 = \{0,0.5,1\}$ introduced in Equation \eqref{data gen} controls the generation of three scenarios:
$\Delta_2 = 0$ corresponds to clusters that differ only in the intercept $\beta_{0k}(t)$, with no influence of covariates;
$\Delta_2 = 1$ corresponds to clusters that differ only in $\beta_k(s,t)$ with no intercept, $\forall k = 1,2,3$;
$\Delta_2 = 0.5$ generates clusters characterized simultaneously by different functional intercepts and regression coefficient functions.

\begin{figure}
    \centering
    \includegraphics[width=0.35\columnwidth]{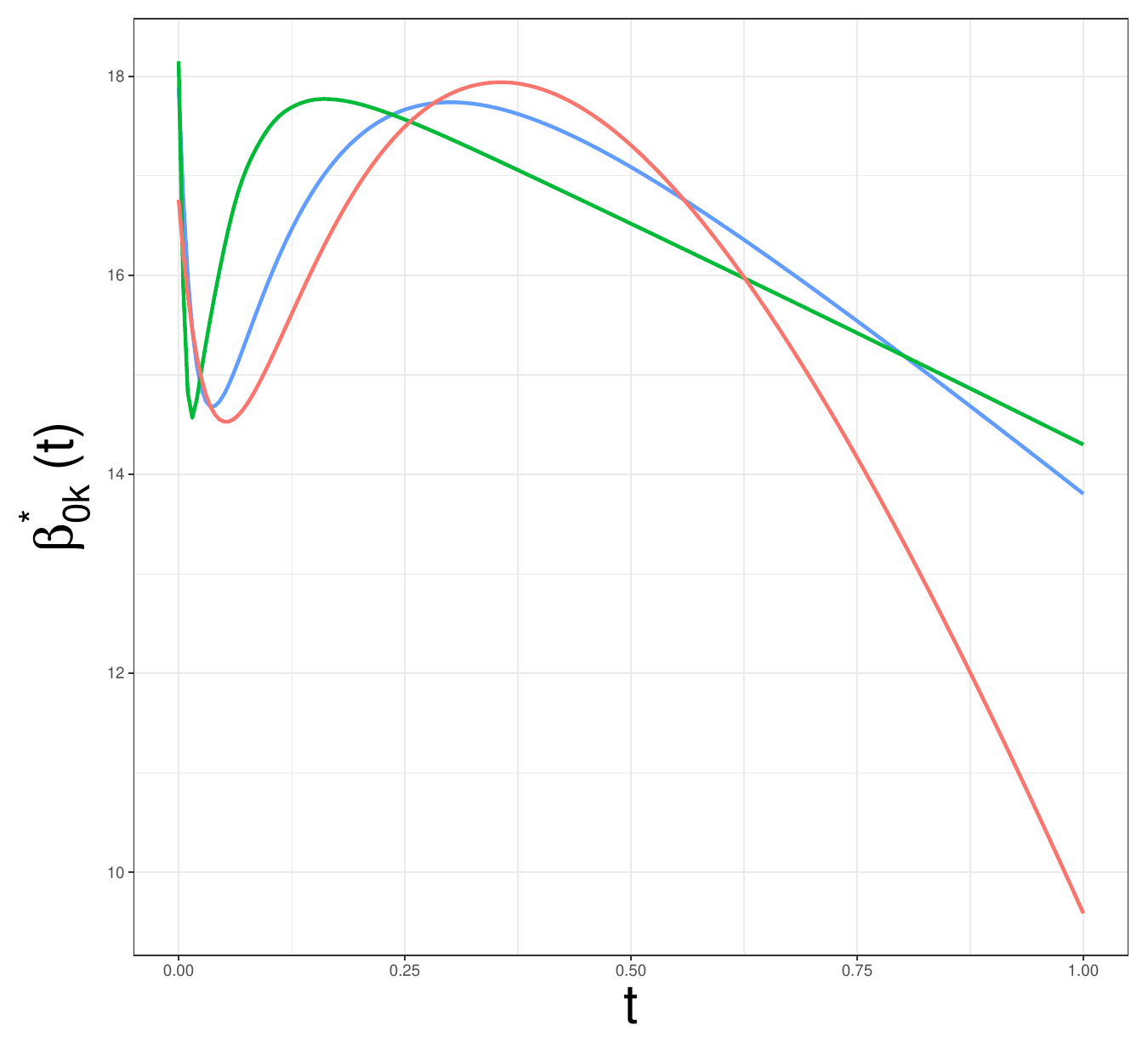}
    \caption{Generated functional intercepts $\beta_{0k}^*(t)$ }
    \label{fig:intercepts}
\end{figure}

\begin{figure}
    \centering
    \includegraphics[width=1\columnwidth]{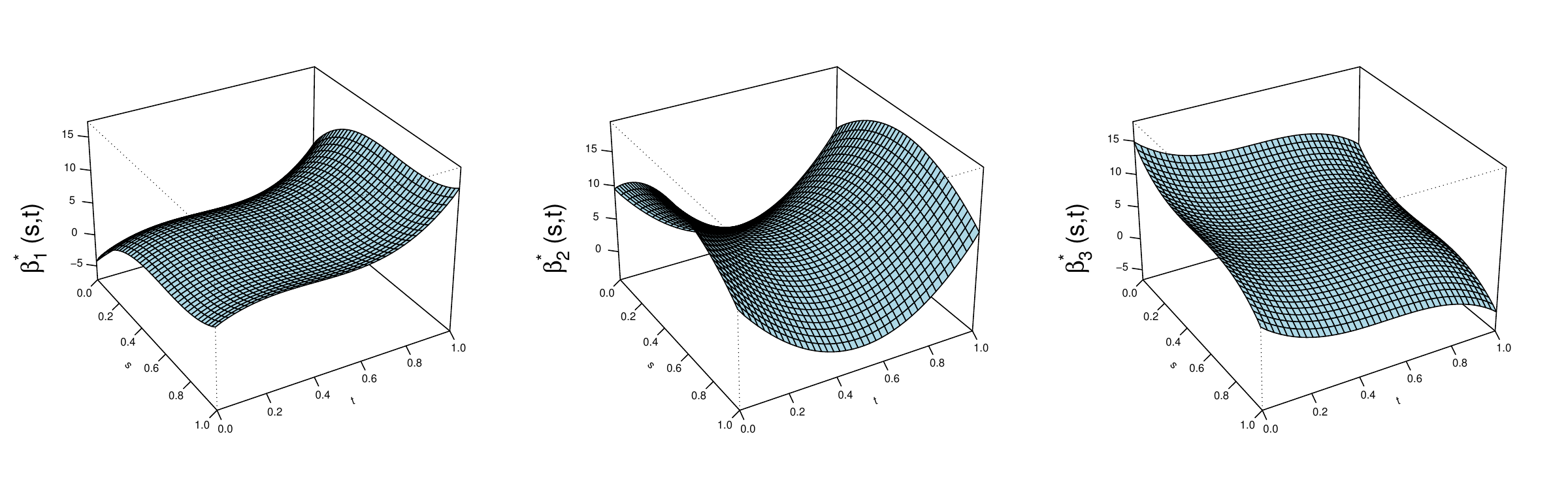}
    \caption{Generated coefficient functions $\beta_k^*(s,t)$}
    \label{fig: beta functions}
\end{figure}

The functional intercepts $\beta_{0k}(t)$ and the regression coefficient functions $\beta_{k}(s,t)$ are obtained as follows
\begin{equation}
\label{}
    \beta_{0k}(t) = (1-\Delta_1)\beta_{01}^*(t) + \Delta_1\beta_{0k}^*(t) \quad t \in \mathcal{T}, k = 2,3,
\end{equation}
\begin{equation}
\label{}
    \beta_{k}(s,t) = (1-\Delta_1)\beta_{1}^*(s,t) + \Delta_1\beta_{k}^*(s,t) \quad t \in \mathcal{T}, k = 2,3,
\end{equation}
where $\beta_{0k}^*(t)$ and $\beta_{k}^*(s,t)$, $k = 1,2,3$, are the different functional intercepts and regression coefficient functions depicted in Figure \ref{fig:intercepts} and Figure \ref{fig: beta functions}, respectively.
The parameter $\Delta_1 = \{0,0.33,0.66,1\}$ controls the dissimilarity between the generated clusters: the larger $\Delta_1$ the more distinct the clusters. Accordingly, $\Delta_1 = 0$ corresponds to a single cluster.
A more detailed description of the data generation process is given in the Supplementary Materials.

Although data are observed through noisy discrete values, each component of the generated observations is obtained by the spline smoothing approach with a roughness penalty in the second derivative, using 80 cubic B-splines, with the penalty parameter chosen via the generalized cross-validation criterion (GCV), as commonly described by \cite{ramsay}.

The purpose of the simulation is to assess the performance of FMRCC in identifying any deviation from the IC heterogeneous population, as defined in Section \ref{monitoring}, in the presence of changes in the mean function of $\Tilde{Y}$ conditional on $\Tilde{X}$, $\E \left(\Tilde{Y}(t)|\Tilde{X}\right)$.
From Equation \eqref{FLMk regression fun} the shifts in $\E \left(\Tilde{Y}(t)|\Tilde{X}\right)$ can result from the changes in $\beta_{0k}(t)$ and $\beta_k(s,t)$. However, the latter, in addition, can also affect the variability of the functional regression residuals. Because we are interested in the performance of FMRCC in identifying mean function shifts in response, given that the variability of residuals is assumed constant, only shifts caused by changes in $\beta_{0k}(t)$ are considered.
Both linear and quadratic types of the shift are considered to represent a change in the slope or curvature of the FLM.

\subsection{Simulation results and discussion}
The proposed FMRCC is compared to three different profile monitoring methods: (a) the functional regression control chart (FRCC) of \cite{centofanti2021bfunctional}; (b) a functional control chart (FCC), which monitors the scores coming from the functional principal component decomposition of the standardized response $Y$ via Hotelling's $T^2$ and SPE control charts; (c) the FCC applied after a functional clustering step, based on the works of \cite{jacques2013funclust} and \cite{grasso2017phase}, referred to as CLUST.
More specifically, the latter consists of a model-based clustering step on the principal component scores of $Y$ and the use of Hotelling's $T^2$ and SPE control chart on each cluster separately. 
The BIC and the maximum a posterior principle (MAP) are adopted to choose the number of clusters and assign the observations to each group, respectively.

\begin{figure}
    \centering
    \includegraphics[width=1\columnwidth]{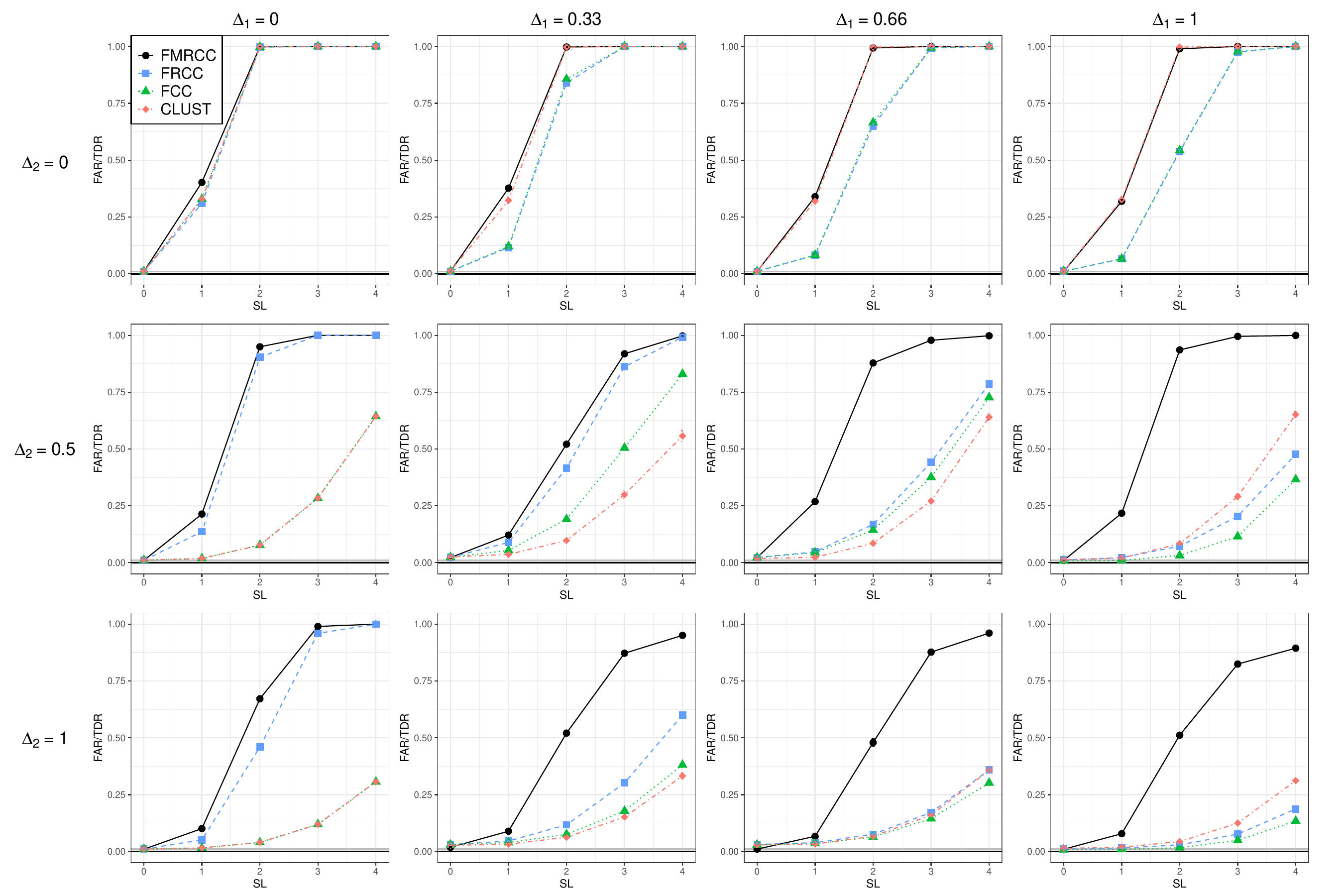}
    \caption{Mean FAR (Severity = 0) or TDR achieved in Phase II by FMRCC, FRCC, FCC, and CLUST for each combination of $\Delta_1$ and $\Delta_2$ for the shift type linear as a function of the severity level.}
    \label{fig:linear}
\end{figure}

Four severity levels of the mean shift (see the Supplementary Materials) are explored.
For each combination of $\Delta_1$, $\Delta_2$, shift type and severity level, 100 simulation runs are performed. 
According to Section \ref{monitoring}, in each run, a training set and a tuning set are independently simulated from IC patterns with 400 and 1000 observations for each cluster, respectively, to mimic a Phase I sample.
The number of principal components $L$ and $M$ retained for $X$ and $Y$ is chosen so that the fraction of variance explained (FVE) is at least $95\%$.

\begin{figure}
    \centering
    \includegraphics[width=1\columnwidth]{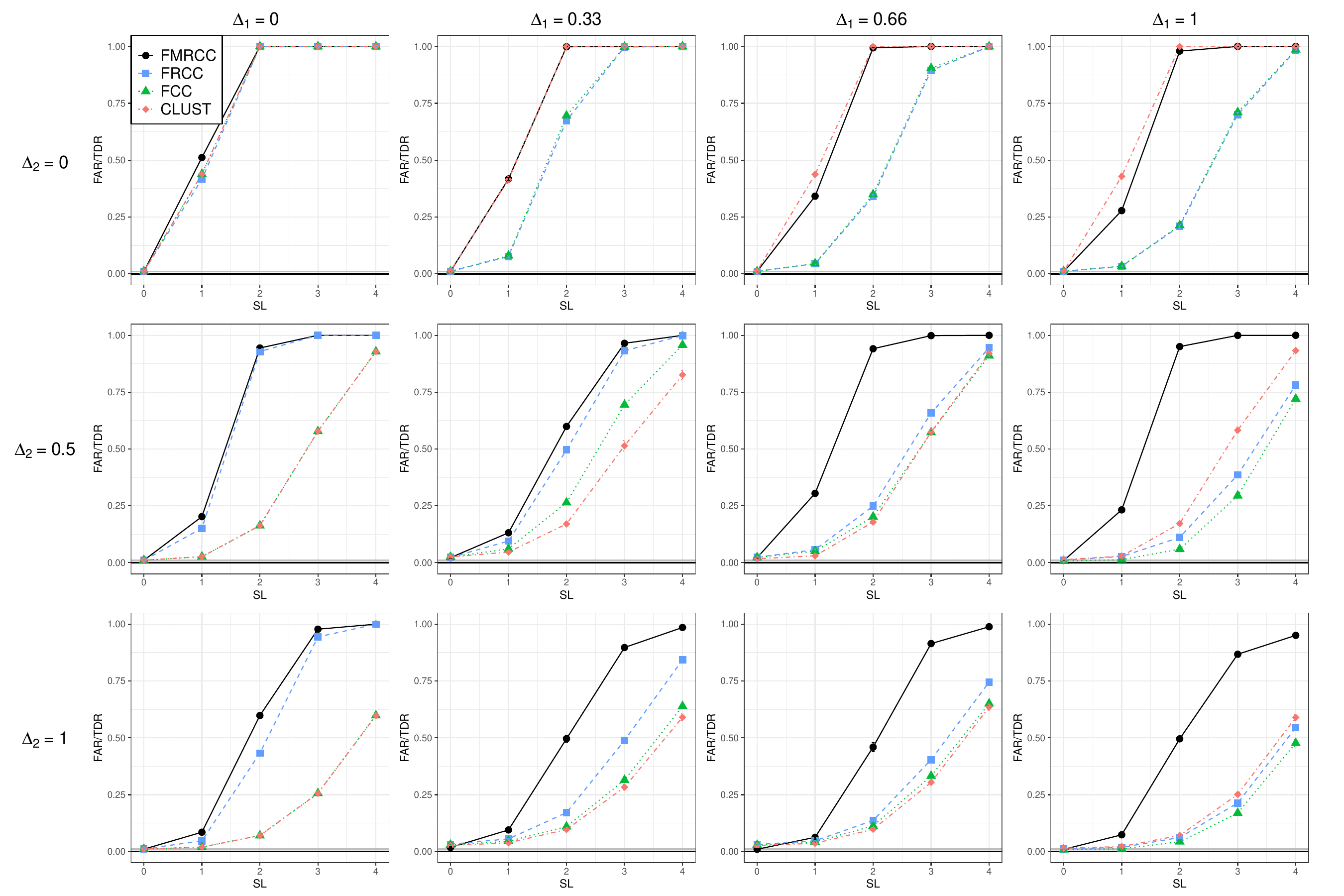}
    \caption{Mean FAR (Severity = 0) or TDR achieved in Phase II by FMRCC, FRCC, FCC, and CLUST for each combination of $\Delta_1$ and $\Delta_2$ for the shift type quadratic as a function of the severity level.}
    \label{fig:quadratic}
\end{figure}

In Phase II, a testing set of 3000 OC observations is randomly generated, and the control chart performance is evaluated by means of the mean true detection rate (TDR) and the mean false alarm rate (FAR), which are estimated as the average proportion, over the simulation runs, of points that fall outside the control limits, whilst the process is, respectively, OC or IC. The mean
FAR should be as similar as possible to the overall type-I error probability, whereas the mean
TDR should be as large as possible.

The simulation results for the linear shift and the quadratic shift, respectively, are displayed in Figure \ref{fig:linear} and Figure \ref{fig:quadratic}, which show the mean FAR and TDR as a function of the severity level for each combination of $\Delta_1$, $\Delta_2$, and severity level.
For the linear shift, when the clusters differ in the functional intercept $\beta_{0k}(t)$ and there is no functional slope ($\Delta_2 = 0$), the first row of panels in Figure \ref{fig:linear} shows that, as the dissimilarity between the clusters increases, FRCC and FCC decrease their performance, while, as expected, CLUST and FMRCC have stable and similar TDRs.
However, when no clusters are present (corresponding to $\Delta_1 = 0$) FMRCC performs slightly better than FRCC.
When clusters differ in functional intercept and, simultaneously, in regression coefficient functions (corresponding to $\Delta_2 = 0.5$), FMRCC does not detect the correct number of clusters for a value of $\Delta_1 = 0.33$, while it succeeds in detecting the correct number of clusters for a value of $\Delta_1 = 0$, $\Delta_1 = 0.66$ and $\Delta_1 = 1$. As expected, incorporating covariate information is key to the proposed FMRCC performance. In particular, CLUST selects the correct number of clusters for each value of $\Delta_1$, due to the presence of different functional intercepts $\beta_{0k}(t)$.
However, the poor performance of the FRCC, which does not account for the different FLM structures, shows its inability to deal with a heterogeneous population.
A more specific advantage of the proposed FMRCC arises when the clusters differ in regression coefficient functions $\beta_k(s,t)$ only ($\Delta_2 = 1$), while keeping the intercept $\beta_{0k}(s,t) = 0$. In this situation, depicted in the third row of Figure \ref{fig:linear}, FMRCC largely outperforms all competing methods and succeeds in detecting the correct number of clusters for each value of $\Delta_1$, while CLUST fails since the common functional intercept induces it to prefer only one cluster. FRCC, even worse than the previous case, is inadequate for capturing the different regression structures.
For the quadratic shift, the simulation results shown in Figure \ref{fig:quadratic} are analogous to the case shown in Figure \ref{fig:linear}. Minor differences arise in the competing method's capability of detecting a change in the profile pattern curvature compared to a modification in slope. FMRCC still outperforms all competing methods when the regression coefficient functions differ from 0, i.e., $\Delta_2 \neq 0$, while remaining competitive when $\Delta_2 = 0$.

\section{Case Study}
\label{sec_real}

\begin{figure}
    \centering
    \includegraphics[width=1\columnwidth]{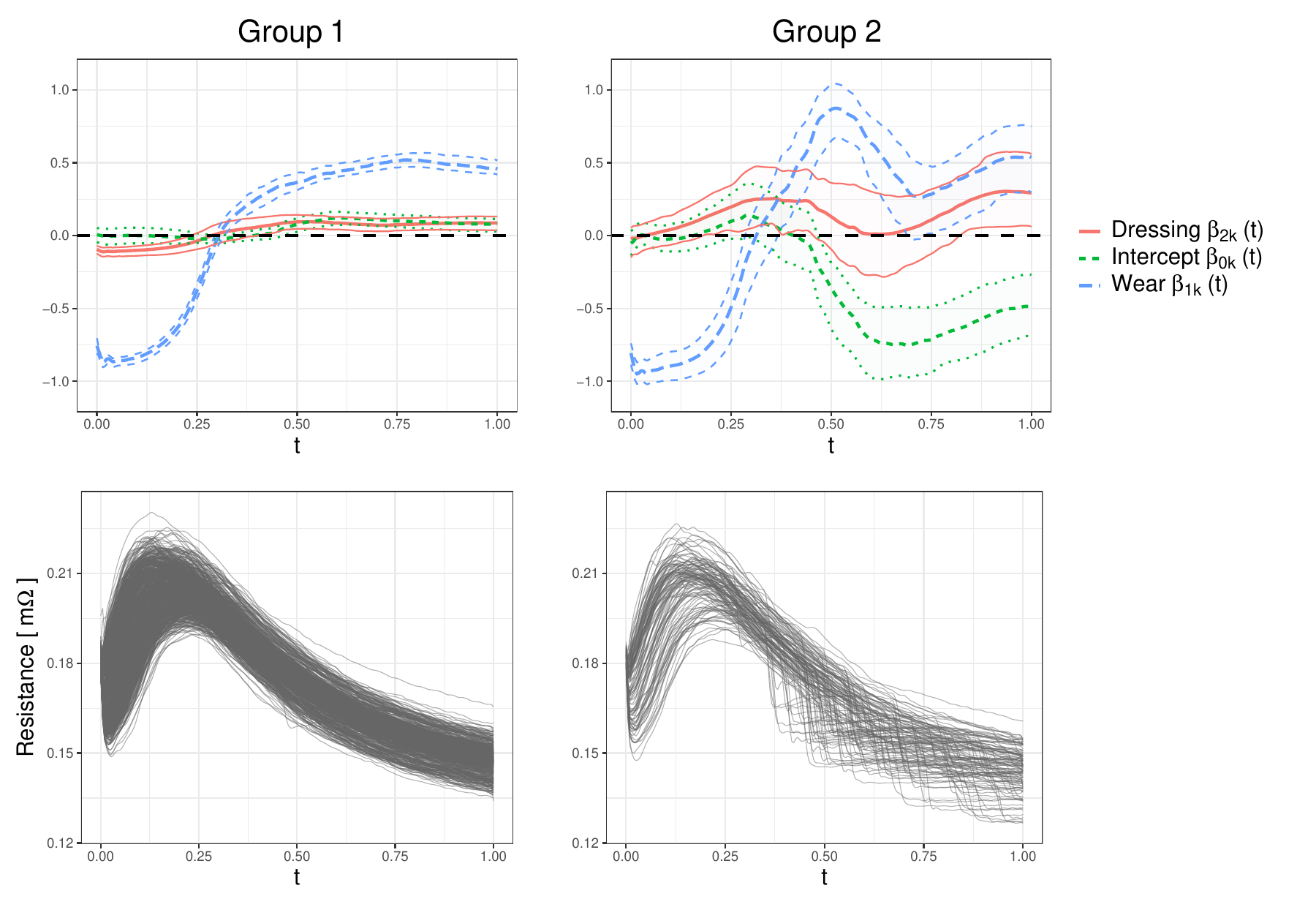}
    \caption{Top panels: Estimated coefficient functions of the two groups detected by the FMRCC along with 95\% pointwise confidence intervals (CI). Bottom panels: The corresponding DRCs assigned by the MAP rule to each group.}
    \label{fig:coef_clust}
\end{figure}

The practical applicability of the FMRCC is demonstrated through the case study, already mentioned in the introduction, on the SPM of an RSW process in the automotive industry through the observations of the DRC, which is known to be an informative proxy of the final quality of a welded spot.
Several factors are known to affect the behavior of DRCs, such as the expulsion of molten material from the weld and the wear of the electrode tips \citep{capezza2024adaptive}.
In particular, the expulsion phenomenon is associated with a notable resistance drop and is usually due to an excessive current and/or inadequate clamping force in the welding spot \citep{valaee2020criterion}.
However, recent findings from downstream ultrasound inspections, routinely used to certify the actual quality of spot welds on a very low percentage of car bodies, suggest that many spot welds showing signs of molten material expulsion are not necessarily defective. In fact, under certain conditions, expulsion does not always compromise the final quality of the spot weld. Therefore, the corresponding DRCs, despite exhibiting heterogeneous behavior, should be included in the Phase I sample, leading to multimodal IC data.
Along the same line, electrode tip wear, which is known to cause changes in electrical, thermal, and mechanical conditions at the electrode tip and metal sheets interfaces \citep{manladan2017review},
is counteracted by periodic electrode tip dressing to avoid inadequate final quality of welded joints. Therefore, its consequences on DRC behavior, already depicted in Figure \ref{fig: Wear Dressing effect}, may still be representative, up to a certain level, of the IC state and lead to another IC mode in the 1802 IC DRCs of the RSW dataset considered in the Phase I sample. The remaining 37 OC DRCs were instead used as Phase II observations.
\begin{figure}
    \centering
    \includegraphics[width=0.5\columnwidth]{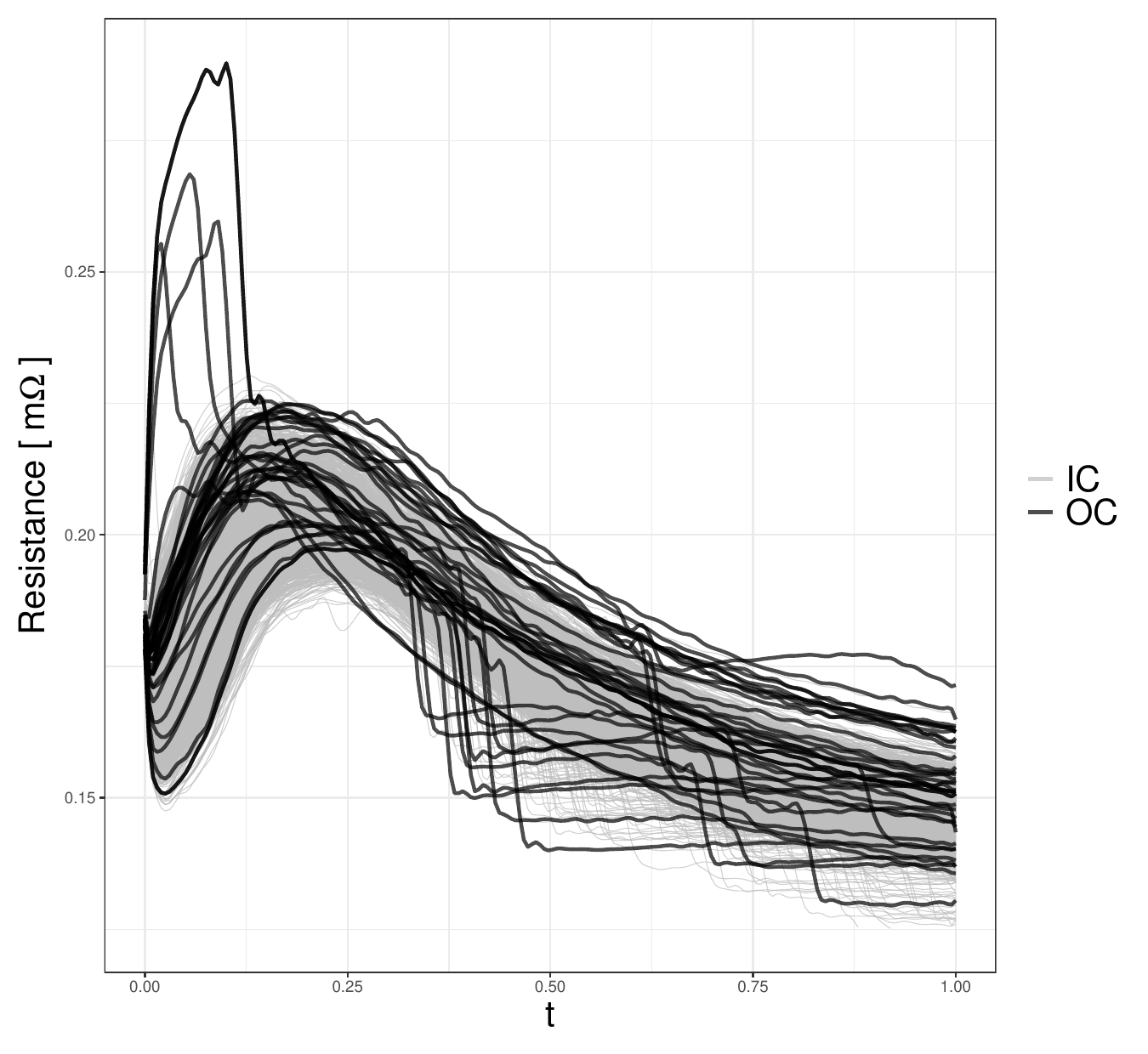}
    \caption{37 OC DRCs used in Phase II on the IC Phase I DRCs.}
    \label{fig:OC}
\end{figure}
For illustrative purposes, in Figure \ref{fig:OC}, the OC DRCs (black line) are superimposed on the IC DRCs (grey line).
It is worth remarking that, although the FMRCC framework is developed in the more general case of univariate functional response and multivariate functional predictors, in this case study the two predictors are scalar and will not be subject to dimension reduction.
As described in Section \ref{methodology}, the Phase I sample is partitioned into training and tuning sets by randomly selecting 901 observations without remittance for the former and the remaining 901 observations for the latter.
Five components of the functional response $Y$ are retained to account for at least 95\% of the total variance explained.
Two mutually exclusive groups with different regression structures are suggested by BIC.
The estimates of the regression functions with the 95\% pointwise confidence intervals shown in the top panels of Figure \ref{fig:coef_clust} are obtained based on 100 bootstrap samples via the non-parametric bootstrap approach of \cite{yao2011functional}, which includes a label-switching strategy to avoid non-identifiability of component labels.
The corresponding DRCs assigned by the MAP rule are shown in the bottom panels of \ref{fig:coef_clust}.
The intercept term of group 2 reveals a drop in DRCs, which clearly corresponds to spot welds affected by the expulsion phenomenon.
The wear effect appears to be clearly mapped by a resistance decrease in the first portion of the DRC domain in both clusters, while the dressing effect manifests itself by an overall modest increase of the resistance in the DRC, with a slight decrease in a small initial domain portion.
\begin{figure}
    \centering
    \includegraphics[width=1\columnwidth]{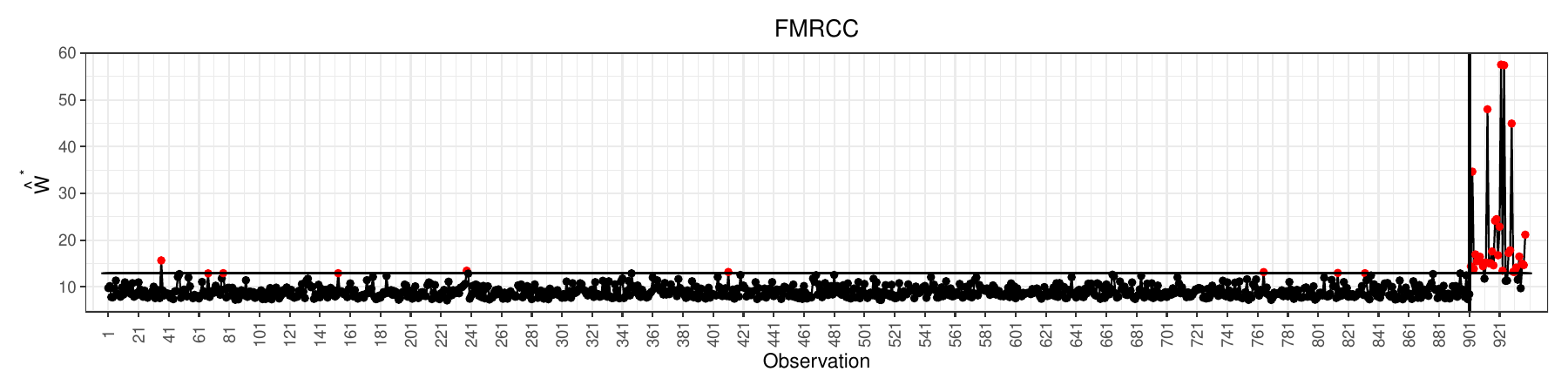}
    \caption{The proposed FMRCC applied to the RSW dataset. The vertical line separates the monitoring statistic calculated for the tuning set on the left and the Phase II observations on the right, while the control limit is shown as a horizontal line.}
    \label{fig:cc}
\end{figure}

Figure \ref{fig:cc} shows the proposed FMRCC method applied to the RSW dataset. The vertical line separates the monitoring statistic  $\hat{W}^*$ calculated for the tuning set, on the left, and the Phase II observations, on the right, while the control limit is represented by the horizontal line. 
This figure shows that 32 over 37 OC DRCs are correctly detected, resulting in an estimated TDR, denoted by $\widehat{TDR}$, of 0.864. 
This value is reported in Table \ref{tab:TDR case study} along with that achieved by the competing methods on the RSW dataset, with the same training and tuning sets and Phase II observations.
Finally, to quantify the uncertainty of the $\widehat{TDR}$, Table \ref{tab:TDR case study} reports the bootstrap 95\% confidence intervals (CI) of the empirical bootstrap distribution \citep{efron1986bootstrap} of $\widehat{TDR}$ for the FMRCC and the competing methods presented in Section \ref{sec_sim}.
The CI achieved by the FMRCC is above and non-overlapping those of all competing approaches and confirms its superior performance also in this real scenario, where the capability of simultaneously accounting for the variability explained by the covariates and the heterogeneous structure of the population is decisive.

\begin{table}[]
\centering
\resizebox{0.55\linewidth}{!}{
\begin{tabular}{p{3cm}p{0.5cm}p{3cm}c}
\toprule
      &  & $\widehat{TDR}$ &  CI  \\ 
      \cline{3-4} \\[-0.7em]
FMRCC &  & 0.864      & {[}0.756,0.960{]} \\  \\[-0.7em]
FRCC  &  & 0.486       & {[}0.310,0.581{]} \\  \\[-0.7em]
FCC   &  & 0.486       & {[}0.323,0.608{]} \\  \\[-0.7em]
CLUST &  & 0.621       & {[}0.472,0.729{]} \\ 
\bottomrule
\end{tabular}}
\caption{Estimated TDR values, denoted as $\widehat{TDR}$, on the Phase II observations and the corresponding bootstrap 95\% confidence interval (CI) for each monitoring method.}
\label{tab:TDR case study}
\end{table}

\section{Conclusions}
\label{sec_conclusions}
A new framework, referred to as functional mixture regression control chart (FMRCC), is proposed for statistical process monitoring (SPM) of a functional quality characteristic linked to functional and/or scalar covariates by possibly different functional linear models (FLM).
The FMRCC is the first profile monitoring framework that is able, by means of a likelihood-raio test (LRT), to jointly enhance the SPM performance by exploiting additional information on covariates and allowing the subjects to obey different FLMs, thus simultaneously modeling the variability explained by covariates and the hidden heterogeneity of the population structure.

From an extensive Monte Carlo simulation, the FMRCC demonstrated, in the function-on-function regression setting, its superiority in identifying mean shifts in the response over three competing methods suggested by the multimode profile monitoring literature.
The practical applicability of the proposed method is illustrated through the case study that motivated this research on the SPM of a resistance spot welding (RSW) process in the automotive industry where heterogeneity of the functional quality characteristic, the dynamic resistance curve (DRC), is modeled through additional information on the wear of the electrode tips and the number of tip dressings the electrode has been subject to. In the case study, the FMRCC confirms that it outperforms all competitors with a more prompt ability to identify the out-of-control state of the RSW process.

Future research can be addressed to extend the proposed framework to nonlinear functional models or to combine it with profile registration techniques and handle profile observations with different domains. In addition, time-weighted monitoring strategies, such as cumulative sum (CUSUM) and exponentially weighted moving average (EWMA), could be implemented to increase the reactivity of the control chart when small and persistent shifts from the in-control population are of interest.

\section*{Supplementary Materials}
Supplement A contains additional details about the data generation process in the simulation study, omitted from the main manuscript for brevity.

\section*{Funding}
The research activity of A. Lepore and F. Centofanti were carried out within the MICS (Made in Italy – Circular and Sustainable) Extended Partnership and received funding from the European Union Next-GenerationEU (PIANO NAZIONALE DI RIPRESA E RESILIENZA (PNRR) – MISSIONE 4 COMPONENTE 2, INVESTIMENTO 1.3 – D.D. 1551.11-10-2022, PE00000004). The research activity of B. Palumbo was carried out within the MOST - Sustainable Mobility National Research Center and received funding from the European Union Next-GenerationEU (PIANO NAZIONALE DI RIPRESA E RESILIENZA (PNRR) – MISSIONE 4 COMPONENTE 2, INVESTIMENTO 1.4 – D.D. 1033.17-06-2022, CN00000023). This work reflects only the authors’ views and opinions, neither the European Union nor the European Commission can be considered responsible for them.

\bibliographystyle{apalike}
\setlength{\bibsep}{5pt plus 0.2ex}
{\small
\spacingset{1}
\bibliography{ref}
}

\end{document}

\def\spacingset#1{\renewcommand{\baselinestretch}%
{#1}\small\normalsize} \spacingset{1}
\newcommand{\specialcell}[2][c]{%
  \begin{tabular}[#1]{@{}c@{}}#2\end{tabular}}

\if0\blind
{
  
\title{Supplementary Materials to ``Functional Mixture Regression Control Chart''}

\author[]{Christian Capezza}
\author[]{Fabio Centofanti}
\author[]{Davide Forcina}
\author[]{Antonio Lepore}
\author[]{Biagio Palumbo\thanks{Corresponding author. e-mail: \texttt{biagio.palumbo@unina.it}}}

\affil[]{Department of Industrial Engineering, University of Naples Federico II, Piazzale Tecchio 80, 80125, Naples, Italy}

\setcounter{Maxaffil}{0}
\renewcommand\Affilfont{\itshape\small}
\date{}
\maketitle
} \fi

\if1\blind
{
  \bigskip
  \bigskip
  \bigskip
  \begin{center}
    {\LARGE\bf Supplementary Materials to ``Robust Multivariate Functional Control Charts''}
\end{center}
  \medskip
} \fi

\spacingset{1.45} % DON'T change the spacing!
\appendix
\numberwithin{equation}{section}

\section{Details on Data Generation in the Simulation Study}
\label{sec_appA}
The data are generated according to the FLM
\begin{equation}
\label{data gen}
    \Tilde{Y}(t) = (1-\Delta_2) \beta_{0k}(t) + \int_{\mathcal{S}} \Delta_2(\beta_k(s,t))^T \Tilde{X}(s)ds + \varepsilon(t), \quad t \in \mathcal{T}, k = 1,\dots,K,
\end{equation}
where the compact domains $\mathcal{S}$ and $\mathcal{T}$ are set, without loss of generality, equal to $\left[0,1\right]$ and the number of covariates $p$ and clusters $K$ are set equal to 1 and 3, respectively.
The functional covariates $\Tilde{X}(s)$ are generated by the eigenfunction set $\lbrace\psi_l\rbrace_{l=1,2,\dots}$, as in \cite{centofanti2021bfunctional}, by considering the correlation function $G$ through the following steps.
\begin{enumerate}
    \item Set $G(s,t) = G(z)$, where $z = |s-t|$, as the \textit{powered exponential} correlation function \citep{stein1999interpolation}. 
    The general form of the correlation function and parameter used are listed in Table \ref{ta_corf}. 

    \item Calculate the eigenvalues $\lambda_{l}$ and the corresponding eigenfunctions $\psi_l$, $l=1,2,\dots,L^*$, of $G$, where $L^*$ is set equal to 50.

    \item Let $\Tilde{X}$ as
    \begin{equation}
    \Tilde{X}=\sum_{l=1}^{L^*}\xi_l^{\Tilde{X}}\psi_l^{\Tilde{X}},
    \end{equation}
    with $\xi_{L^{*}}^{\Tilde{X}}=\left(\xi^{\Tilde{X}}_1,\dots,\xi^{\Tilde{X}}_{L^{*}}\right)^{T}$ generated by means of a multivariate normal distribution with mean 0 and covariance $\Cov\left(\bm{\xi}_{L^{*}}^{\Tilde{X}}\right)=\diag\left(\lambda_1,\dots,\lambda_{L^{*}}\right)$. 

\end{enumerate}
Covariate generation is made by means of the R package \texttt{funcharts} of \cite{capezza2023funcharts}.

Inspired by the coefficient functions generation process proposed in \cite{centofanti2022smooth}, the regression coefficient functions $\beta_k^*(s,t)$, depicted in Figure 3, are obtained through the following reference model for $k=1,2,3$
\begin{equation}
    \label{beta fun formula}
    \beta_k^*(s,t) = \left(\frac{t-0.5}{c}\right)^3 +
    \left(\frac{s-0.5}{d}\right)^3 +
    \left(\frac{t-0.5}{b}\right)^2 -
    \left(\frac{s-0.5}{a}\right)^2 + e, \quad s,t \in [0,1],
\end{equation}
where the parameters $c,d,b,a,e$ are real numbers depicted in Table \ref{tab:param beta}.

\begin{table}
\caption{Powered exponential correlation function and parameter for covariates generation in the simulation study.}
\label{ta_corf}
\centering
\resizebox{0.4\textwidth}{!}{
\begin{tabular}{ccc}
\toprule
&$\rho$&$\nu$\\
\midrule
$P\left(z\right)= \exp{- \left(\frac{|z|}{\rho}\right)^\nu}$&1&0.5\\[.35cm]
\bottomrule
\end{tabular}}
\end{table}

\begin{table}[]
\centering
\caption{Parameters $a$,$b$,$c$,$d$,$e$ used for the functional coefficient functions generation.}
\label{tab:param beta}
\begin{tabular}{lllllll}
\hline
               &  & $a$ & $b$  & $c$  & $d$  & $e$  \\ \cline{1-1} \cline{3-7} 
$\beta_1^*(s,t)$ &  & 0.3 & 0.3  & 0.3  & 0.3  & 5   \\ 
$\beta_2^*(s,t)$ &  & 0.2 & 0.15 & 0.9  & 0.9  & -5  \\  
$\beta_3^*(s,t)$ &  & 0.9 & 0.9  & -0.3 & -0.3 & 5   \\ \hline
\end{tabular}
\end{table}

Furthermore, the functional intercepts $\beta_{0k}^*(t)$, depicted in Figure 2, are obtained through the following reference model for $k=1,2,3$
\begin{multline}
    \beta_{0k}^*(t) = f + 0.3117\exp\left[-371.4u_k(t)\right] +0.5284\{1 - \exp\left[gu_k(t)\right]\} \\ -423.3\{ 1 + \tanh \left[-hu_k(t)+0.1715\right]\}\quad t\in\mathcal{T},
\end{multline}
where the parameters $f$, $g$ and $h$ are real numbers depicted in Table \ref{tab:param int} and $u_k(t)$ is defined as follows
\begin{equation}
        u_k(t) = \frac{t-min(t)}{max(t)-min(t)}(m-0.0045)+0.0045 \quad t\in\mathcal{T},
\end{equation}
with $m = 0.15 , 0.4 , 0.08$ for $k = 1,2,3$ respectively.
Note that the mean functions $\beta_{0k}^*(t)$ are generated to resemble a typical DRC through the phenomenological model for the RSW process presented in \cite{schwab2012improving}.
\begin{table}[]
\centering
\caption{Parameters $f$, $g$ and $h$ used for the functional intercepts generation.}
\label{tab:param int}
\begin{tabular}{lllll}
\hline
               &  & $f$    & $g$   & $h$      \\  \cline{3-5} 
$\beta_{01}^*(t)$ &  & 0.2074 & 0.8217   & 26.15  \\ 
$\beta_{02}^*(t)$ &  & 0.187  & 0.2      & 27  \\  
$\beta_{03}^*(t)$ &  & 0.3    & 4        & 24 \\ \hline
\end{tabular}
\end{table}

The parameter $\Delta_1 = \{0, 0.33, 0.66, 1\}$ represents a measure of dissimilarity between the generated clusters according to the equations
\begin{equation}
\label{}
    \beta_{0k}(t) = (1-\Delta_1)\beta_{01}^*(t) + \Delta_1\beta_{0k}^*(t) \quad t \in \mathcal{T}, k = 2,3
\end{equation}
\begin{equation}
\label{}
    \beta_{k}(s,t) = (1-\Delta_1)\beta_{1}^*(s,t) + \Delta_1\beta_{k}^*(s,t) \quad t \in \mathcal{T}, k = 2,3.
\end{equation}
A higher value of $\Delta_1$ corresponds to more distinct clusters, whereas a value of $\Delta_1 = 0$ corresponds to the absence of clusters.

The parameter $\Delta_2 = \{0, 0.5, 1\}$ weighs the functional intercept term and the regression coefficient function term according to Equation \eqref{data gen}.
In particular, a value of $\Delta_2 = 0$ corresponds to clusters that differ only in $\beta_{0k}(t)$, while $\beta_k(s,t) = 0 \quad \forall k = 1,2,3$. On the other hand, a value of $\Delta_2 = 1$ corresponds to clusters that differ only in $\beta_k(s,t)$, while $\beta_{0k}(t) = 0 \quad \forall k = 1,2,3$. Finally, a value of $\Delta_2 = 0.5$ indicates the presence of clusters characterized simultaneously by different functional intercepts and regression coefficient functions.

\begin{figure}
    \centering
    \includegraphics[width=1\columnwidth]{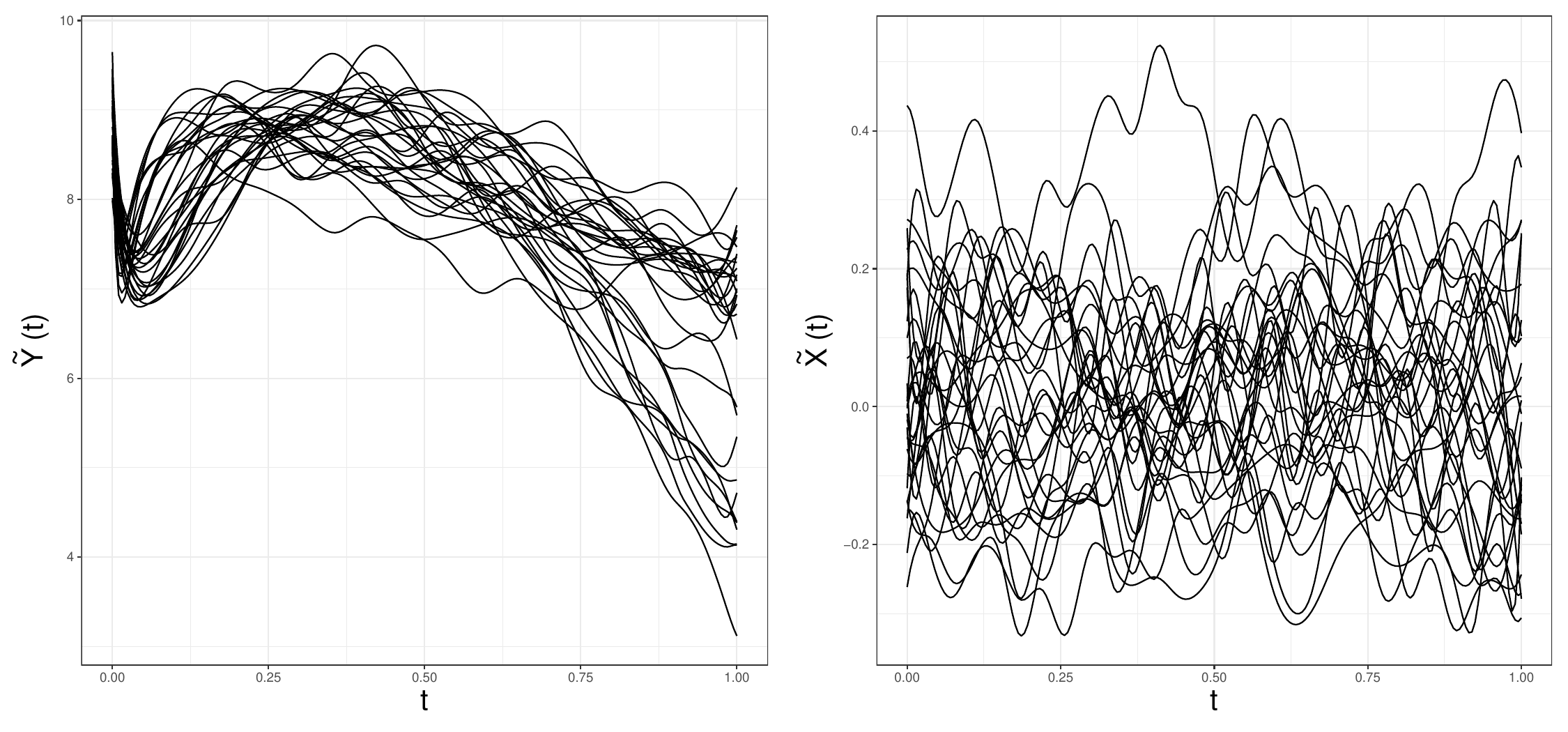}
    \caption{Sample of 30 IC randomly generated $\Tilde{Y}$ and $\Tilde{X}$ functions corresponding to $\Delta_1 = 1$ and $\Delta_2 = 0.5$}
    \label{fig:IC functions}
\end{figure}

Moreover, to have noisy realizations of the response, a functional error term $k\varepsilon(t) = \sum_{i=1}^{20}e_i\psi_i^e(t)$ is generated, where $e_i$ are independent realizations of a standard normal random variable and $\psi_i^e(t)$ are cubic B-splines with evenly spaced knot sequence. The constant $k$ is chosen to have a signal-to-noise ratio (SNR) defined as \cite{stadler2010ℓ} equal to 10.
Finally, it is assumed that the generated data were observed discretely at 500 equally spaced time points throughout the domain $\left[0,1\right]$.

For illustrative purposes, a sample of 30 IC randomly generated $\Tilde{Y}$ and $\Tilde{X}$ functions corresponding to $\Delta_1 = 1$ and $\Delta_2 = 0.5$ are shown in Figure \ref{fig:IC functions}.

Without loss of generality, the Phase II out-of-control functional profiles are generated from the first cluster using the model in Equation \eqref{data gen} with an additional shift term
$\delta(t)$ defined as follows
\begin{equation}
    \label{intercept shift}
    \delta(t) = \delta_1t^2 + \delta_2t, \quad t \in [0,1].
\end{equation}
The real numbers $\delta_1$ and $\delta_2$ are set equal to positive severities as reported in Table \ref{tab:shift types}, where two different shift types are considered.
The terms $q_l$ and $q_q$ are set equal to $1.2$ and $1.6$, respectively, and $s$ represents five severity levels equal to $0,0.375,0.75,1.25,1.5$.

\begin{table}[]
\centering
\caption{Shift types according to Equation \ref{intercept shift}}
\label{tab:shift types}
\begin{tabular}{p{3cm}p{3cm}p{1cm}}
\hline
Shift     & $\delta_1$ & $\delta_1$ \\ \hline
\\[-0.7em]
Linear    & 0 & $s\cdot q_l$  \\
Quadratic & $s\cdot q_q$ & 0  \\
\\[-1em]
\hline
\end{tabular}
\end{table}

\noindent The linear shift is representative of a change in the slope of the profile pattern, whereas the quadratic shift represents curvature modification.

\bibliographystyle{chicago}
{\small
\bibliography{ref}}